# A New Virtual Oscillator based Grid-forming Controller with Decoupled Control Over Individual Phases and Improved Performance of Unbalanced Fault Ride-through

Ritwik Ghosh, Narsa Reddy Tummuru, *Senior Member, IEEE*, and Bharat Singh Rajpuohit, *Senior Member, IEEE*

*Abstract*— Virtual Oscillator (VO) control is the latest and promising control technique for grid-forming and grid-supporting inverters. VO Controllers (VOCs) provide time-domain synchronization with a connected electrical network. At the same time, a VOC can incorporate additional nested control loops to meet the system-level requirements such as fault ride-through capability. However, existing VOCs have two limitations. Firstly, the existing VOCs do not have decoupled control over individual phases. As a result, the performance of the existing VOCs is not satisfactory in presence of unbalanced grid voltages. Secondly, the fault ride-through performance of the existing VOCs under unbalanced faults is not satisfactory. The power operation at a healthy phase is badly affected by a faulty phase. This paper has introduced a modified VO based system-level grid-forming controller to overcome the limitations mentioned above. Systematic development of the proposed control architecture with analytical reasoning is presented. Simulation studies and hardware experiments are conducted for validation.

## I. INTRODUCTION

Virtual Oscillator (VO) based Grid Forming (GFM) control strategy has gone through several major developments in recent years. The concept of decentralized communication-free control of inverters using non-linear limit cycle oscillators is first proposed in [1]–[3]. It is followed by crucial component level development and analysis. Virtual Oscillator based Controller (VOC) provides all the steady-state functionalities of a conventional droop controller [4]. At the same time, the dynamic performance of VOC is better than the conventional droop controller [5]–[7]. Different oscillator models such as dead zone, Van-der-pol, modified Van-der-pol, Andronov-Hopf, Dispatchable Virtual Oscillator (d-VOC) are presented in [8]–[13]. Dead-zone and Van-der-Pol oscillators are not suitable for three-phase operations because they are not able to incorporate more than one input as feedback [14]. Dead-zone and Van-der-Pol oscillators are also unable to provide decoupled control over real and reactive power without additional control loops [11]. Van-der-Pol oscillators can achieve faster dynamics only at an expense of higher harmonic content at the output voltage [9]. A modification in the non-linear current source of the Van-der-pol oscillator is proposed in [10] to reduce the third harmonic component from the output voltage. Andronov-Hopf oscillator is proved to be the best fit for grid-forming controllers by overcoming the mentioned limitations of Dead-zone and Van-der-Pol oscillators [11]. The dispatchable VOC (d-VOC) and an Andronov-Hopf VOC have a similar fundamental form [11].

The component-level development is followed by the development of important techniques which help the system-level implementation of VOC. VOC with additional control loops is used for grid-connected operation in [15], [16]. Single VOC is used for islanded and grid-connected operation with seamless transition using hierarchical and unified control structures in [17] and [18] respectively. The Virtual Impedance technique is integrated into VOC for selective harmonic current rejection and over-current protection in [19] and [20] respectively.

The outline of the system-level implementation of VO-based GFM controllers can be divided into two broad categories. In the first category as shown in Fig. 1. (a) one virtual oscillator is used for generating the reference voltage for the inverter. Other cascaded control loops are added for different objectives such as limiting harmonic components from the current output, limiting current amplitude during a fault [19], [20].

The second category is more recent and advanced. In the second category as shown in Fig. 1. (b) a dispatchable VOC (d-VOC) is used with nested voltage and current control loops [21], [22]. The d-VOC provides continuous reference voltage and phase angle information. The phase information is used for park transformation in the nested voltage and current control loops. The nested controller tracks the reference voltage, generated by the d-VOC in synchronous reference frame. The nested controller also can provide additional functionalities such as protection against large grid side transients.

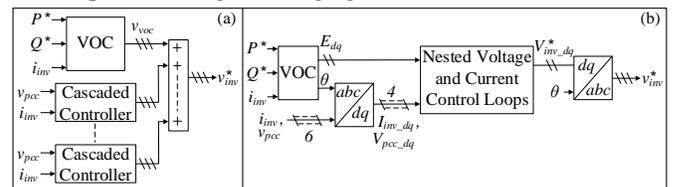

Fig. 1. The existing control structures for system-level implementation of VOC

However, the existing literature has not considered two important issues related to the system-level implementation of VOC.

(i) Decoupled Control over individual phases: Without the ability of decoupled control over individual phases, a VOC fails to perform in presence of unbalanced grid voltage [23]. It is important to mention that short duration (3cycle – 1min) voltage unbalance of 2-15% and a steady-state voltage unbalance of 0.5-5% are very usual in electrical power systems [24]. Among all the power quality events the unbalance in grid voltages is countable for approximately 32.3%. [25]. The voltage unbalance has adverse effects on the loads and the reliability of the electrical network [26], [27]. It is expected from the distributed energy sources that they will support the grid with reactive power in presence of unbalanced voltage sags [28]–[30]. In our last work [23] we have presented the Symmetrical Component based VOC (S-VOC) to address the issue. However, S-VOC alone can not provide all the system-level functionalities such as capacitor voltage control, over current protection. In this paper, a complete outline of the system-level implementation of S-VOC as shown in Fig. 2 is presented. The S-VOC is used to generate continuous reference voltages and phase information for individual phases. Then each phase is equipped with one nested voltage controller and one nested current controller to achieve decoupled control. The nested voltage control loop is also equipped with an anti-windup controller to provide over-current protection against large grid transients.

(ii) Riding through unbalanced low voltage faults: The fault ride-through requirements are becoming strict for distributed renewable energy sources in recent grid codes [31]. Grid forming inverters should stay connected with the electrical network and inject reactive power to the network under balanced and unbalanced faults [31], [32]. The performance of any inverter controller under an unbalanced fault is very important [33], [34]. During unbalanced fault conditions, the controller is expected to limit the output current and inject reactive power at faulty phases [35], [36]. At the same time, the control over the healthy phases should be affected as minimum as possible [35], [36]. The existing VOC based grid-forming controller as presented in [21], [22] cannot provide satisfactory performance under unbalanced fault conditions. A simulation study is presented in Section III where an existing VOC based grid-forming controller is used to ride through an unbalanced fault condition. It is observed that the current at the faulty phase is restricted. However, the control over the healthy phases is affected severely by the currents of the faulty phases. The healthy phases stop injecting power without any valid reason. This paper has addressed the issue by introducing a modified feedback system in the fault ride-through controller.

The rest of the paper is organized as follows. In Section II the overall systematic development of the proposed controller is presented. In Section III the performance of the existing VOC based grid-forming controller under an unbalanced fault is presented using a simulation study. Section IV has presented a modified fault ride-through technique that improves the performance of the VOC based grid-forming controller under unbalanced faults. Simulation studies and hardware experiments that validate the performance of the proposed S-VOC-based grid forming controller are presented in Section V and Section VI respectively. Finally, the paper is concluded in Section VII.

## II. THE PROPOSED CONTROL STRUCTURE

The schematic diagram of the proposed S-VOC based system-level grid-forming controller is shown in Fig. 2. The proposed controller consists of three main functional building blocks which are an S-VOC, reference frame transformers, and nested voltage and current control loops. The detailed schematic diagram of the S-VOC is presented in Fig. 3. The S-VOC provides continuous synchronization with the positive, negative, and zero sequences of a connected electrical network. The output of the S-VOC is three reference voltages for the individual phases of a grid-forming inverter. The S-VOC also provides the lagging orthogonal components of the three reference voltages. The references for individual phases are independent of each other as shown in Fig. 4. The individual references are used to transform the instantaneous parameters ($v_{gx}$, $i_{invx}$) to synchronous reference frame-based parameters ($V_{gx}$, $I_{invx}$) as shown in Fig. 4. The nested voltage and current control loops produce the reference voltages for the inverter in synchronous reference frame. The detailed schematic diagram of the nested voltage and current controllers is shown in Fig. 5.

### A. The model of the S-VOC

The S-VOC consists of three Andronov-Hopf oscillators which generate the positive, negative, and zero sequence reference voltages. As required by each sequence the primary Andronov-Hopf oscillator model [11] is modified. Individual reference voltage for each phase of a three-phase inverter is generated by adding the outputs of the three oscillators. The three output phase currents of the three-phase inverter are converted to instantaneous symmetrical components and are used as feedback to the S-VOC.

The oscillator which generates the positive sequence reference voltages is presented as

$$\dot{v}_{+\alpha} = \frac{\xi}{k_v^2}(2V_n^2 - ||v_{+\alpha\beta}||^2)v_{+\alpha} + \omega_n v_{+\beta} + \frac{k_v k_i}{C}(i_{+\beta f}) \quad (1)$$

$$\dot{v}_{+\beta} = \frac{\xi}{k_v^2}(2V_n^2 - ||v_{+\alpha\beta}||^2)v_{+\beta} - \omega_n v_{+\alpha} - \frac{k_v k_i}{C}(i_{+\alpha f}) \quad (2)$$

where, $V_n$, $\omega_n$, $i_{+\alpha f}$, $i_{+\beta f}$ are the rms nominal voltage, nominal frequency, feedback currents respectively.

The oscillator which generates the negative sequence reference voltages is presented as

$$\dot{v}_{-\alpha 1} = \frac{\xi}{k_v^2}(2V_n^2 - ||v_{-\alpha\beta 1}||^2)v_{-\alpha 1} + \omega_n v_{-\beta 1} + \frac{k_v k_i}{C}(-i_{-\beta f}) \quad (3)$$

$$\dot{v}_{-\beta 1} = \frac{\xi}{k_v^2}(2V_n^2 - ||v_{-\alpha\beta 1}||^2)v_{-\beta 1} - \omega_n v_{-\alpha 1} - \frac{k_v k_i}{C}(i_{-\alpha f}) \quad (4)$$

$$v_{-\alpha} = (||v_{-\alpha\beta 1}|| - \sqrt{2}\,V_n)\frac{v_{-\alpha 1}}{||v_{-\alpha\beta 1}||} \quad (5)$$

$$v_{-\beta} = -(||v_{-\alpha\beta 1}|| - \sqrt{2}\,V_n)\frac{v_{-\beta 1}}{||v_{-\alpha\beta 1}||} \quad (6)$$

where, $\omega_n$, $i_{-\alpha f}$, $i_{-\beta f}$ are the nominal frequency, feedback currents respectively. The rms nominal voltage of the oscillator is zero.

The oscillator which generates the zero sequence reference voltages is presented as

$$\dot{v}_{0\alpha 1} = \frac{\xi}{k_v^2}(2V_n^2 - ||v_{0\alpha\beta 1}||^2)v_{0\alpha 1} + \omega_n v_{0\beta 1} + \frac{k_v k_i}{C}(i_{0\beta f}) \quad (7)$$

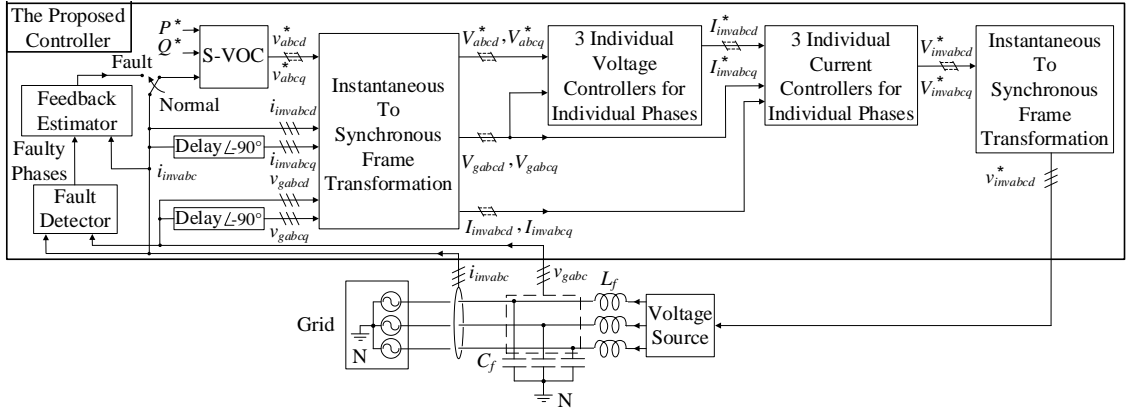

Fig. 2. The schematic diagram of the proposed S-VOC based grid-forming controller

$$\dot{v}_{0\beta1} = \frac{\xi}{k_v^2}(2V_n^2 - ||v_{0\alpha\beta1}||^2)v_{0\beta1} - \omega_n v_{0\alpha1} - \frac{k_v k_i}{C}(i_{0\alpha f}) \tag{8}$$

$$v_{0\alpha} = (||v_{0\alpha\beta1}|| - \sqrt{2}\,V_n)\frac{v_{0\alpha1}}{||v_{0\alpha\beta1}||} \tag{9}$$

$$v_{0\beta} = (||v_{0\alpha\beta1}|| - \sqrt{2}\,V_n)\frac{v_{0\beta1}}{||v_{0\alpha\beta1}||} \tag{10}$$

where, $\omega_n$, $i_{0\alpha f}$, $i_{0\beta f}$ are the nominal frequency, feedback currents respectively. The rms nominal voltage of the oscillator is zero.

The positive, negative, and zero sequence reference voltages are given by

$$\begin{bmatrix} v_{+ad} \\ v_{+bd} \\ v_{+cd} \end{bmatrix} = \begin{bmatrix} 1 & 0 \\ -\frac{1}{2} & \frac{\sqrt{3}}{2} \\ -\frac{1}{2} & -\frac{\sqrt{3}}{2} \end{bmatrix} \begin{bmatrix} v_{+\alpha} \\ v_{+\beta} \end{bmatrix}; \begin{bmatrix} v_{+aq} \\ v_{+bq} \\ v_{+cq} \end{bmatrix} = \begin{bmatrix} 1 & 0 \\ -\frac{1}{2} & \frac{\sqrt{3}}{2} \\ -\frac{1}{2} & -\frac{\sqrt{3}}{2} \end{bmatrix} \begin{bmatrix} -v_{+\beta} \\ v_{+\alpha} \end{bmatrix} \tag{11}$$

$$\begin{bmatrix} v_{-ad} \\ v_{-bd} \\ v_{-cd} \end{bmatrix} = \begin{bmatrix} 1 & 0 \\ -\frac{1}{2} & \frac{\sqrt{3}}{2} \\ -\frac{1}{2} & -\frac{\sqrt{3}}{2} \end{bmatrix} \begin{bmatrix} v_{-\alpha} \\ v_{-\beta} \end{bmatrix}; \begin{bmatrix} v_{-aq} \\ v_{-bq} \\ v_{-cq} \end{bmatrix} = \begin{bmatrix} 1 & 0 \\ -\frac{1}{2} & \frac{\sqrt{3}}{2} \\ -\frac{1}{2} & -\frac{\sqrt{3}}{2} \end{bmatrix} \begin{bmatrix} v_{-\beta} \\ -v_{-\alpha} \end{bmatrix} \tag{12}$$

$$\begin{bmatrix} v_{0ad} \\ v_{0bd} \\ v_{0cd} \end{bmatrix} = \begin{bmatrix} 1 \\ 1 \\ 1 \end{bmatrix} [v_{0\alpha}]; \begin{bmatrix} v_{0aq} \\ v_{0bq} \\ v_{0cq} \end{bmatrix} = \begin{bmatrix} 1 \\ 1 \\ 1 \end{bmatrix} [v_{0\beta}] \tag{13}$$

The individual reference voltage for each phase of the three-phase inverter is given by

$$v_{ad}^* = v_{+ad} + v_{-ad} + v_{0ad};\ v_{aq}^* = v_{+aq} + v_{-aq} + v_{0q} \tag{14}$$
$$v_{bd}^* = v_{+bd} + v_{-bd} + v_{0bd};\ v_{bq}^* = v_{+bq} + v_{-bq} + v_{0q} \tag{15}$$
$$v_{cd}^* = v_{+cd} + v_{-cd} + v_{0cd};\ v_{cq}^* = v_{+cq} + v_{-cq} + v_{0q} \tag{16}$$

where, $v_{abcq}^*$ are orthogonal to $v_{abcd}^*$ as shown in Fig. 4.

The reference phase currents for the S-VOC are given by

$$\begin{bmatrix} i_{xd}^* \\ i_{xq}^* \end{bmatrix} = \frac{2}{(v_{xd}^*)^2 + (v_{xq}^*)^2} \begin{bmatrix} v_{xd}^* & v_{xq}^* \\ v_{xq}^* & -v_{xd}^* \end{bmatrix} \begin{bmatrix} P_x^* \\ Q_x^* \end{bmatrix}; x \in a,b,c \tag{17}$$

where, $P_a^*, P_b^*, P_c^*$ and $Q_a^*, Q_b^*, Q_c^*$ are the power setpoints. The current feedback to the S-VOC is derived as

$$i_{af} = i_{a\_inv} - i_a^*;\ i_{bf} = i_{b\_inv} - i_b^*;\ i_{cf} = i_{c\_inv} - i_c^* \tag{18}$$

where, $i_{abc\_inv}$ are the output phase currents of the three-phase inverter.

The current feedback is converted to symmetrical components and is applied to the three oscillators as

$$A = e^{\frac{2}{3}\pi j};\ A^2 = e^{\frac{4}{3}\pi j} \tag{19}$$

$$\begin{bmatrix} i_{+af} \\ i_{+bf} \\ i_{+cf} \end{bmatrix} = \frac{1}{3}\begin{bmatrix} 1 & A & A^2 \\ A^2 & 1 & A \\ A & A^2 & 1 \end{bmatrix}\begin{bmatrix} i_{af} \\ i_{bf} \\ i_{cf} \end{bmatrix} \tag{20}$$

$$\begin{bmatrix} i_{-af} \\ i_{-bf} \\ i_{-cf} \end{bmatrix} = \frac{1}{3}\begin{bmatrix} 1 & A^2 & A \\ A & 1 & A^2 \\ A^2 & A & 1 \end{bmatrix}\begin{bmatrix} i_{af} \\ i_{bf} \\ i_{cf} \end{bmatrix} \tag{21}$$

$$i_{0f} = \frac{1}{3}(i_{af} + i_{bf} + i_{cf}) \tag{22}$$

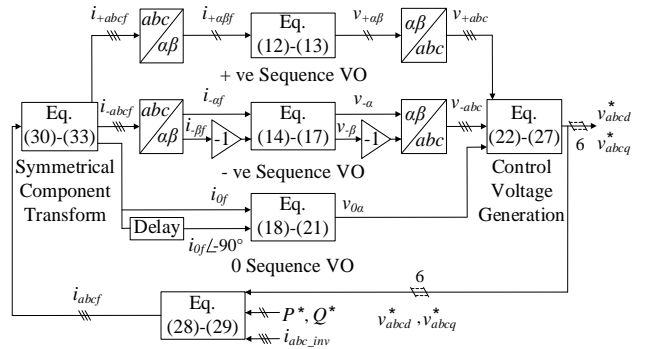

Fig. 3. Symmetrical Component based Virtual Oscillator Controller (S-VOC)

*B. Transformation of instantaneous to synchronous reference frame*

The S-VOC provide three reference voltages i.e $v_{ad}^*$, $v_{bd}^*$, and $v_{cd}^*$ for three individual phases. The instantaneous parameters of a phase are converted to synchronous reference frame-based parameters using the reference voltage of the same individual phase as shown in Fig. 4. The instantaneous parameters are denoted by small letters. The symmetrical reference frame-based parameters are denoted by capital letters.

The transformation of the output voltages of the S-VOC is derived as

$$V_{xd}^* = \sqrt{(v_{xd}^*)^2 + (v_{xq}^*)^2}\ ;\ V_{xq}^* = 0\ ;(x \in a,b,c) \tag{23}$$

The transformation of the voltages of the PCC and the output currents of the inverter is derived as

$$\hat{v}_{xd}^* = \frac{v_{xd}^*}{V_{xd}^*}\ ;\ \hat{v}_{xq}^* = \frac{v_{xq}^*}{V_{xd}^*}\ ;(x \in a,b,c) \tag{24}$$

$$\begin{bmatrix} V_{gxd} \\ V_{gxq} \end{bmatrix} = \begin{bmatrix} \hat{v}_{xd}^* & \hat{v}_{xq}^* \\ \hat{v}_{xq}^* & -\hat{v}_{xd}^* \end{bmatrix}\begin{bmatrix} v_{gxd} \\ v_{gxq} \end{bmatrix};(x \in a,b,c) \tag{25}$$

$$\begin{bmatrix} I_{invxd} \\ I_{invxq} \end{bmatrix} = \begin{bmatrix} \hat{v}_{xd}^* & \hat{v}_{xq}^* \\ \hat{v}_{xq}^* & -\hat{v}_{xd}^* \end{bmatrix}\begin{bmatrix} i_{invxd} \\ i_{invxq} \end{bmatrix};(x \in a,b,c) \tag{26}$$

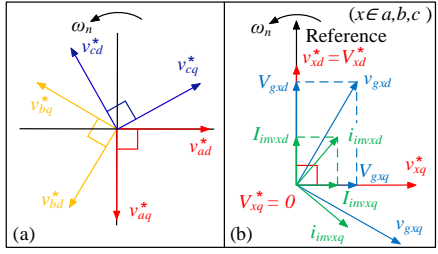

Fig. 4. (a) Output voltage vectors of the S-VOC (b) Instantaneous to synchronous reference frame transformation

*C. The nested voltage and current control loops*

Each phase is equipped with an individual nested control loop. The block diagram of one nested control loop which consists of an inner current controller and outer voltage controller is shown in Fig. 5. The Anti-windup technique is used in the voltage controller to limit current output during a fault condition. The nested controller produces the individual reference voltage for each phase of the three-phase inverter in synchronous reference frame.

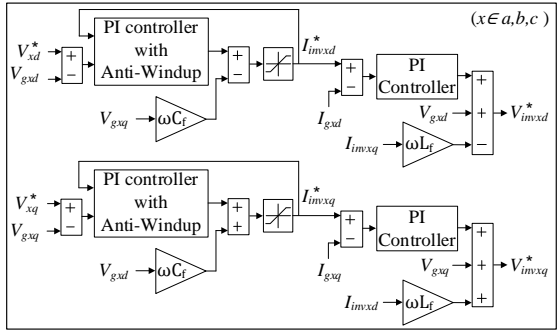

Fig. 5. The nested control loop

*D. Synchronous to instantaneous frame transformation*

The reference voltages for the inverter in instantaneous reference frame is derived as

$$[v^*_{invx}] = [\hat{v}^*_{xd} \quad \hat{v}^*_{xq}] \begin{bmatrix} V^*_{invxd} \\ V^*_{invxq} \end{bmatrix}; \; (x \in a,b,c) \quad (27)$$

## III. SIMULATION STUDY: EXISTING VOC BASED GRID-FORMING CONTROLLER UNDER UNBALANCED FAULT

This section presents a simulation study where an existing VOC is used during an unbalanced low voltage fault. The simulation study helps to analyze the problem faced by the existing controller during any unbalanced fault.

*A. Simulation study: Existing VOC in presence of an unbalanced fault*

The schematic diagram of the system, used in the simulation study is shown in Fig. 6. The specification of the system is given in Table I. An existing VO-based grid-forming system-level controller as presented in [21], [22] is used in the simulation study. The nested voltage loop is equipped with an anti-windup controller to limit the fault current. However, the following simulation study has shown that the anti-windup controller alone can not provide a satisfactory performance during an unbalanced fault.

From t = 0.5 s to t = 5.5 s a low voltage fault condition is created at phase-a of PCC from the grid side as shown in Fig. 6. The voltage amplitude of the PCC of phase-a is lowered to 10% of the nominal voltage from the grid side. The voltage amplitude of phase-b and phase-c is kept constant at the nominal voltage from the grid side. The current and power output of the individual phases of the voltage source during the unbalanced fault condition is shown in Fig. 7. The anti-windup controller successfully limits the output current of phase-a under the predefined limit, $I_{max}$. The voltage source also supports phase-a of the PCC with reactive power. However, the current and power output at the healthy phases (phase-b and phase-c) are completely undesirable during the fault condition. The voltage source stop feeding power to the grid at the healthy phases (phase-b and phase-c) because of the low voltage fault at the other phase (phase-a). This is not a satisfactory performance during an unbalanced fault and it complicates the fault condition further by not supporting the healthy phases properly.

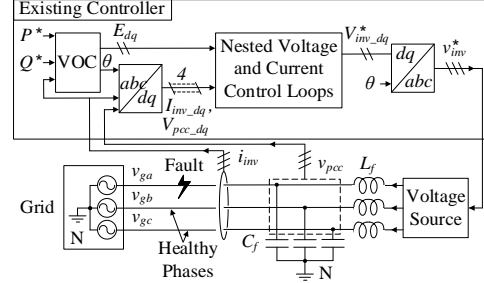

Fig. 6. Existing VO-based grid-forming system-level controller in presence of an unbalanced fault

Table I: Specifications of system used for the simulation study

| Symbol | Description | Value |
|---|---|---|
| $V_{ng}$ | Nominal grid voltage: Phase (rms) | 50 V |
| $\omega_{ng}$ | Nominal frequency | $2\pi 50$ rad/s |
| $L_f$ | Filter inductor | 2 mH/Phase |
| $C_f$ | Filter capacitor | 20 μF/Phase |
| $I_{max}$ | Overcurrent limit: Phase (rms) | 10 A |

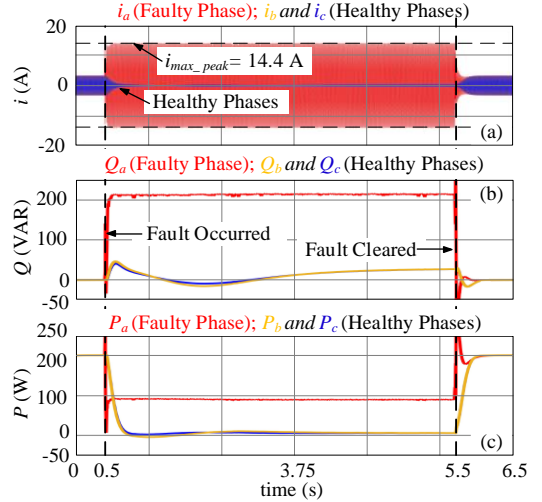

Fig. 7. The existing VOC during unbalanced fault (one phase under fault): (a) individual phase currents (b) reactive power outputs of individual phases (c) active power outputs of individual phases

*B. The reason behind the unsatisfactory performance of the existing VOC during the unbalanced fault*

During any fault condition, the controller enters into the current control mode to stay connected with the network by limiting the output currents. The Virtual Oscillator (VO) block of the controller works as a PLL during the fault condition to provide synchronization with the connected network. The feedback to the VO is the output currents of the voltage source. In normal balanced conditions, the output current is the result of the interaction between the voltages of the PCC ($v_{pcc\_x}; x \in a,b,c$) and the voltages of the VO ($v_{voc\_x}; x \in a,b,c$). In the unbalanced fault condition as shown in Fig. 7, at the faulty phase (phase-a) the current is limited by the anti-windup loop. Once the saturation hits the output current of phase-a, $i_{inv\_a}$ is no longer the result of the interaction between $v_{pcc\_a}$ and $v_{voc\_a}$. The output current of phase-a, $i_{inv\_a}$ becomes the result of the anti-windup controller. At the same time the currents of phase-b and phase-c, ($i_{inv\_b}$ and $i_{inv\_c}$) remains the results of the interaction between ($v_{pcc\_x}; x \in b,c$) and ($v_{voc\_x}; x \in b,c$). This is the reason why the VO fails to provide proper synchronization during the unbalanced fault condition. Due to the improper synchronization, the output powers at the healthy phases become undesirable which makes the situation worse during the fault.

## IV. THE PROPOSED MODIFIED FAULT-RIDE THROUGH TECHNIQUE FOR UNBALANCED FAULTS

The performance of the proposed S-VOC based grid-forming controller is improved during an unbalanced fault condition by modifying the feedback. The modification helps to achieve two objectives simultaneously. The first objective is to limit the current output of the faulty phase and also to support the faulty phase with reactive power. The second objective is to maintain the preset power outputs at the healthy phases.

The objective is fulfilled by considering the healthy phases as the reference to get synchronized with the connected electrical network during an unbalanced fault condition. During an unbalanced fault, the output currents of the healthy phases are used to derive the estimated current feedback for the faulty phases as shown in Fig. 8. The logic is as follows

Step 1: The fault detector as shown in Fig. 2. detects the faulty phases. The sequences of the faulty phases are fed the next block

Step 2: The fault detector activates the 'Feedback Estimator' block to derive feedbacks for unbalanced fault conditions. The feedbacks for the faulty phases are estimated from the output currents of the healthy phases as shown in Fig. 8.

There are three phases, $P_1$, $P_2$, and $P_3$ where the sequence is predefined and given as
$$P_1 = P_2 \angle 120° = P_3 \angle 240° \quad (28)$$
The feedback estimator uses two discrete logic for two possible conditions.

i. One phase under fault: Among three phases one phase i.e. $i^{th}$ phase $P_i$ is under fault. The feedback for the $i^{th} (i \in 1,2,3)$ phase is derived as
$$i_{inv\_ip} = -\sum i_{inv_j}; (j \in 1,2,3), (j \neq i) \quad (29)$$
ii. Two phases under fault: Among three phases one phase i.e. $k^{th}$ ($k \in 1,2,3$) phase is healthy. The feedback for the faulty phases are derived as
$$i_{inv_{(k+n)p}} = i_{inv_k} \angle -(n \times 120°); n \in 1,2 \quad (30)$$
$$i_{inv_{(k-n)p}} = i_{inv_k} \angle (n \times 120°); n \in 1,2 \quad (31)$$
$$(n+k) \in 1,2,3 \quad (32)$$
Step 3: The feedback to the VOC is replaced by the estimated feedback during an unbalanced fault as shown in Fig. 2.

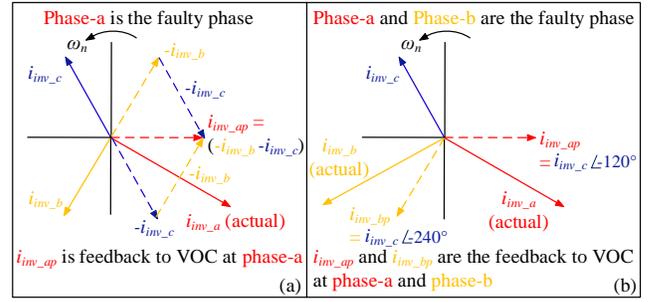

Fig. 8. Feedback modification for unbalanced fault condition: (a) one phase out of three is under fault (b) two phases out of three are under fault

## V. SIMULATION STUDY: THE PROPOSED CONTROLLER

The system used for the simulation study is shown in Fig. 2. The specifications of the system are given in Table I.

### A. Reference power tracking

The reference active power, $P^*$ is set to 600 W initially. At t = 1 s the reference power is changed to 900 W. The current and power output of individual phases are shown in Fig. 9. The controller successfully tracks the reference power.

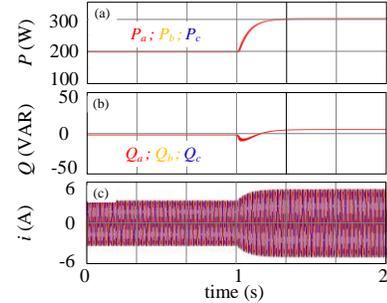

Fig. 9. Reference power tracking by the proposed controller (a) active power outputs of individual phases (b) reactive power outputs of individual phases (c) individual phase currents

### B. Reactive power support during unbalanced voltage sag

The performance of the proposed controller is tested here in presence of 10% sag in one phase of the three-phase network. The voltage amplitude of phase-a at PCC is lowered by 10% from the grid side at t = 0.5. The Current and power output of the grid-forming voltage source is shown in Fig. 10. It is observed that the voltage source supports phase-a with reactive power. The control over phase-b and phase-c remains nearly unaffected.

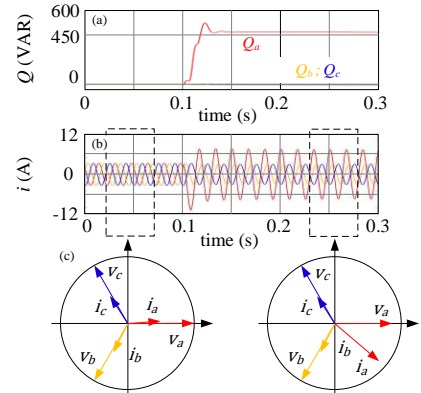

Fig. 10. Reactive power support by the proposed controller during an unbalanced voltage sag: (a) reactive power outputs of individual phases (b) individual phase currents (c) relative phases among individual phase currents of the voltage source and phase voltages of PCC

## C. Unbalanced fault ride-through performance

In this simulation study, the performance of the proposed controller is tested in presence of unbalanced low voltage fault conditions. Two different unbalanced fault conditions are created at the PCC from the grid side. At first, the fault is created at phase-a. The voltage of phase-a at PCC is lowered to 0.1 P.U. from the grid side. The voltage amplitude of phase-b and phase-c is kept constant at the nominal value. Next, the fault is created at phase-b and phase-c. The voltage of phase-b and phase-c at PCC is lowered to 0.1 P.U. from the grid side. The voltage amplitude of phase-a is kept constant at the nominal value. In both the cases, as shown in Fig. 11 and Fig. 12 the proposed controller successfully restricts the output current under $I_{max}$ at the faulty phase. The three-phase voltage sourse also supports PCC at the faulty phases by injecting reactive power. At the same time, the operation at the healthy phase remains nearly unaffected.

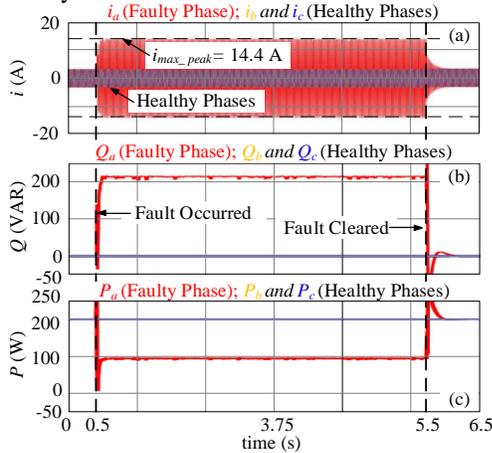

Fig. 11. The proposed S-VOC Grid-forming controller in presence of an unbalanced fault (one phase is under fault): (a) individual phase currents (b) reactive power outputs of individual phases (c) active power outputs of individual phases

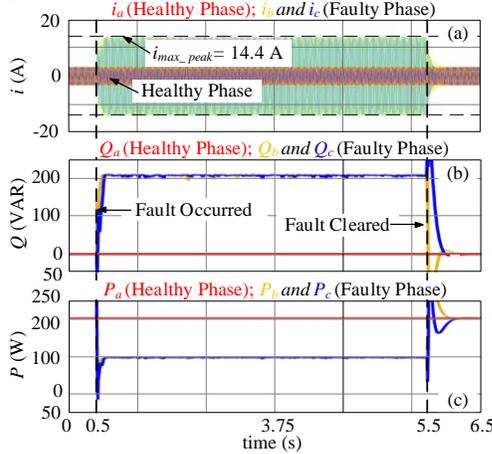

Fig. 12. The proposed S-VOC Grid-forming controller in presence of an unbalanced fault (two phases are under fault): (a) individual phase currents (b) reactive power outputs of individual phases (c) active power outputs of individual phases

## VI. EXPERIMENTAL RESULTS AND DISCUSSIONS

The schematic diagram of the experimental setup is shown in Fig. 13. The picture of the experimental setup is shown in Fig. 14. The specification of the experimental setup is provided in Table II. An OPAL-RT OP4510 is used as the controller.

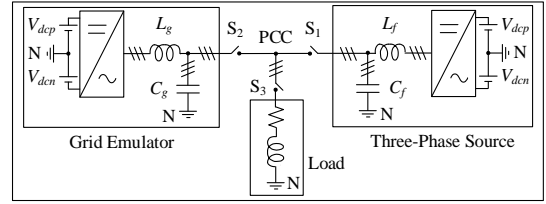

Fig. 13. The schematic diagram of the experimental setup

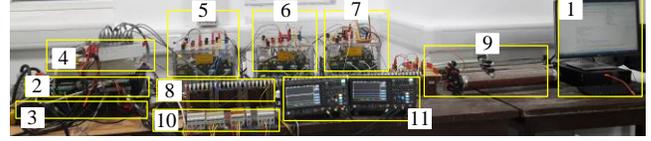

1. Host PC, 2. OPAL-RT, 3. Voltage and Current Sensors, 4. Gate Drivers, 5. 3-φ Inverter, 6. Grid Emulator, 7. dc-sources, 8. LC Filters, 9. RL Load at PCC, 10. Circuit Breakers, 11. MSOs

Fig. 14. The picture of the experimental setup

Table II: Specifications of the experimental setup

| Symbol | Description | Value |
|---|---|---|
| $V_{ng}$ | Nominal voltage: grid emulator | 50 V |
| $\omega_{ng}$ | Nominal frequency: grid emulator | $2\pi 50$ rad/s |
| $L_g$ | Filter inductor: grid emulator | 2 mH/Phase |
| $C_g$ | Filter capacitor: grid emulator | 20 μF/Phase |
| $L_f$ | Filter inductor: sources | 2 mH/Phase |
| $C_f$ | Filter capacitor: sources | 20 μF/Phase |
| $V_{dcp}, V_{dcp}$ | dc-link voltages (split Capacitor) | 100 V |
| Load | RL load at PCC | (5-100 Ω + 2-20 mH)/Phase |
| $S_{rated}$ | Rated kVA | 1 kVA |
| $f_{sw}$ | Switching frequency | 3 kHz |
| $T_s$ | Sampling time of the controller | 50 μs |

### A. Reference power tracking

The voltage profile of the PCC ($v_{PCC}$), output currents of the inverter ($i_{inv}$), and the instantaneous output powers of the inverter ($P_{abc}$ and $Q_{abc}$) are displayed on the MSO using the analog output of the Opal-RT. The voltage profile of the PCC at normal conditions is shown in Fig. 15.

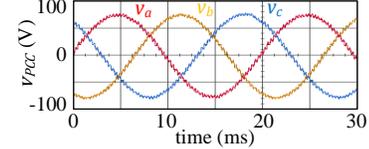

Fig. 15. The voltage profile of the PCC at normal condition

Initially the reference active power, $P^*$ is set to 300 W. As shown in Fig. 16 each phase injects 100 W of active power initially. At t = 100 ms the $P^*$ is changed from 300 W to 900 W. It is seen from Fig. 16 that the controller successfully tracks reference active power. The transient is very close to first-order.

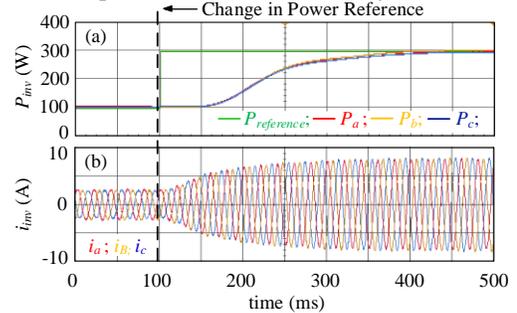

Fig. 16. Reference power tracking by the grid-forming inverter: (a) reference and actual active power outputs of individual phases (b) individual phase currents of the inverter

## B. Reactive power support during unbalanced voltage sags

This experiment validates the reactive power supporting performance of the proposed grid-forming controller in presence of unbalanced voltage sags. The controller is tested under two different conditions of unbalanced voltage sag. The active power reference, $P^*$ is set to 600W. At first, the voltage of phase-a is lowered by 10% of the nominal grid voltage, $V_{gn}$ from the grid emulator. The voltage amplitude of phase-b and phase-c is kept constant at the nominal value. The output reactive powers and currents of the individual phases of the inverter are shown in Fig. 17. It is shown that the inverter starts supporting phase-a of PCC with reactive power in presence of the voltage sag. The operation in phase-b and phase-c remains nearly unaffected.

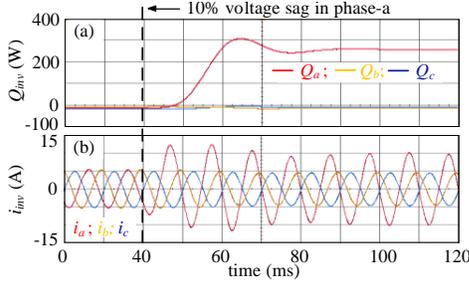

Fig. 17. Reactive power support by the grid-forming inverter in presence of voltage sag in phase-a: (a) reactive power outputs of individual phases (b) individual phase currents of the inverter

Next, the voltage amplitude of phase-b and phase-c is lowered by 10% of the nominal grid voltage, $V_{gn}$ from the grid emulator. The voltage amplitude of phase-a is kept constant at the nominal value. As shown in Fig. 18 the inverter starts injecting reactive power at phase-b and phase-c. The operation at phase-a remains nearly unaffected.

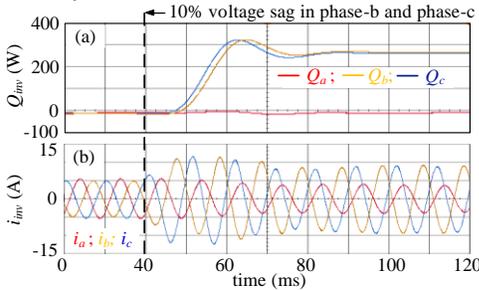

Fig. 18. Reactive power support by the grid-forming inverter in presence of voltage sag in phase-b and phase-c: (a) reactive power outputs of individual phases (b) individual phase currents of the inverter

## C. Unbalanced Fault ride-through performance

The fault ride-through performance of the proposed controller under unbalanced faults is validated by creating two different unbalanced fault conditions. The active power reference, $P^*$ is set to 600W. At first, the voltage amplitude of phase-a of PCC is lowered to 10% of the nominal voltage, $V_{gn}$ from the grid emulator side. The voltage amplitude of phase-b and phase-c is kept constant at the nominal value. The voltage profile of the PCC during the fault is shown in Fig. 19. It is also shown that the inverter successfully limits the output current at phase-a. The inverter starts supporting phase-a of PCC by injecting reactive power. The operation at phase-b and phase-c remains nearly unaffected.

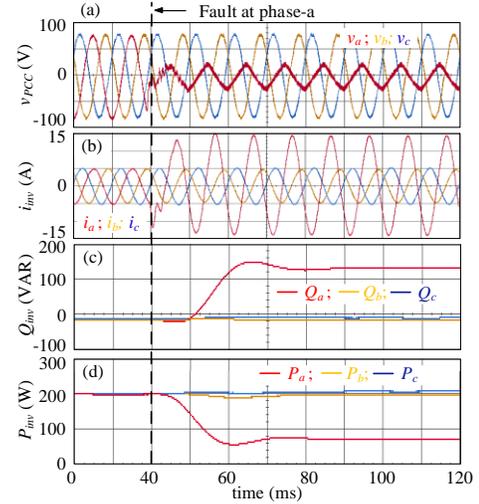

Fig. 19. Fault ride-through of the grid-forming inverter in presence of a fault at phase-a: (a) individual phase voltages of PCC (b) individual phase currents of the inverter (c) reactive power outputs of individual phases (d) active power outputs of individual phases

Next, the voltage amplitude of phase-b and phase-c is lowered to 5% of the nominal voltage, $V_{gn}$. The voltage amplitude of phase-a is kept constant at the nominal value. The voltage profile of the PCC is shown in Fig. 20. It is also seen that the inverter successfully limits the fault current at phase-b and phase-c. The inverter starts supporting phase-b and phase-c with reactive power. The operation at phase-a remains nearly unaffected.

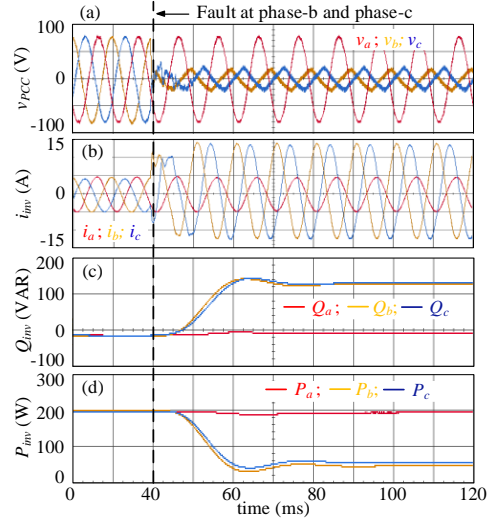

Fig. 20. Fault ride-through of the grid-forming inverter in presence of a fault at phase-b and phase-c: (a) individual phase voltages of PCC (b) individual phase currents of the inverter (c) reactive power outputs of individual phases (d) active power outputs of individual phases

## VII. CONCLUSION

This paper has presented an outline of a Virtual Oscillator (VO) based grid-forming controller. The proposed controller has used the Symmetrical Component based Virtual Oscillator (S-VOC) to achieve decoupled control over individual phases. The decoupled control makes the grid-forming inverters able to perform satisfactorily under unbalanced grid conditions. This paper also introduced a modified fault ride-through strategy for improving the performance of a grid-forming inverter under unbalanced fault conditions. Simulation studies and hardware experiments show that the proposed fault ride-through technique makes a grid-forming inverter able to stay connected with the electrical network during unbalanced fault conditions. The inverter also supports the faulty phases with reactive powers. At the same time, the operation at the healthy phases remains unaffected.